\documentclass[prl,twocolumn,showpacs]{revtex4}
\usepackage{natbib}
\usepackage{epsfig}
\usepackage{graphicx}
\usepackage{psfrag}
\usepackage{amssymb}
\usepackage{times}
 
\begin{document}
 
\title{Comment on `A Wave Traveling over a Hopf Instability Shapes the
Cochlear Tuning Curve'}
\author{R. Stoop}
\author{A. Kern}
\affiliation{Institute of Neuroinformatics, University/ETH Z\"urich,
Winterthurerstr. 190, 8057 Z\"urich}
 
\date{\today}
 
 
\thispagestyle{empty}
\pagestyle{empty}

\begin{abstract}
Since the seminal work by H.L.F. Helmholtz in 1863, to understand
the basic principles of hearing has been a great, but still unresolved, 
challenge for physicists.  Some time ago, it has been pointed out 
(Egu\'{\i}luz et al., Phys. Rev. 
Lett. 84, 5232, 2000) that the generic mathematical 
properties of nonlinear oscillators undergoing a Hopf bifurcation 
account for the salient characteristics of hearing. 
Recently, M.O. Magnasco proposed a model of the cochlea 
(Phys. Rev. Lett. 90, 058101, 2003), which employs
Hopf-type instabilities for cochlear amplification. 
While this model reproduces the input-output behaviour of the 
cochlea to some extent, the generated model responses deviate 
significantly from physiological measurements. 
The reason for the discrepancies between model and experiment
are due to the critical choice of the Hopf control parameter close to 
the bifurcation point ($\mu = 0$). The question whether
the bifurcation parameter has to be chosen critically or subcritically
($\mu < 0$), is central, and has become the subject of a scientific debate.
In this contribution, we argue 
that, for sustained input signals, the control parameter will assume a
subcritical value. This leads to model results that are in
close agreement with reported experimental data. 
\end{abstract}

\maketitle

Egu\'{\i}luz et al. 
\cite{AA:Eguiluz_etal2000} have put forward
the idea that the salient properties of hearing could be described by a
Hopf oscillator system. Their paper, however, left open the actual
detailed implementation of this concept within known properties of the
cochlea. In a recent contribution \cite{AA:Magnasco2003}, M.O. Magnasco 
attempts to close  this gap by
complementing the Hopf system with a model of a wave that interacts
with the Hopf instabilities as it travels along the
cochlea. His results corroborate the generic validity of the
Hopf-approach by
producing responses that are much closer to measured cochlea data 
\cite{AA:Eguiluz_etal2000, AA:Magnasco2003} if compared with the pure
Hopf instabilities 
\cite{AA:Eguiluz_etal2000}. 

In \cite{AA:Magnasco2003}, the Hopf system is considered 
tuned exactly at the bifurcation
point $\mu=0$.  While this approach 
reduces the computation time, it seriously modifies the obtained
cochlear response. 
For the following, we will concentrate on the modifications imposed on
the frequency response curves, 
as they respond more sensitively to modeling details if compared 
to tuning curves.

The first consequence is that nonlinear responses  ($R\sim F^{1/3}$) are
obtained for sustained
sound inputs, irrespective of the stimulus size. This is in
contradiction to the
constant-gain regime for small stimuli that has previously 
been outlined (experimentally: bottom curves in Fig. 2 (a) of 
\cite{AA:Magnasco2003}; theoretically: see Ref.
\cite{AA:Eguiluz_etal2000}). 

Secondly, the amplifier bandwidth becomes exceedingly small
($\Gamma_{3dB} \sim F^{2/3}$) \cite{AA:Eguiluz_etal2000}, 
which leads to
abrupt  discontinuous model responses. It is true that in nonstationary
driving conditions, $\mu$ may
approach or even cross zero (e.g. in order to reduce the response time
for incoming follow-up signals). 
For sustained signals, however, $\mu$ will assume  a non-zero equilibrium value
\cite{AA:Camalet_etal2000}. As the transition points
between linear and compressive regime, as well as the amplifier
bandwidths, are smooth functions of $\mu<0$,
this can consistently be exploited for obtaining realistic response
curves \cite{AlbertDiss}. We emphasize that the filter-based passive model of
\cite{AA:Magnasco2003} 
already mimicks longitudinal BM coupling. Therefore, the negligence
of the longitudinal BM stiffness should not be considered as a
substantial cause for the obtained 
response curve discontinuities. Moreover, also contributions from
additional paths with  $\omega \neq \omega_{ch}$ are unable to
remove this effect, but instead will merely result in a smoothening of
the sharp transition at the response peak. Simulations with models that
include passive BM couplings \cite{unpublished} clearly show that the
defect persists as long as $\mu=0$ is chosen. 

Thirdly, we argue that the choice $\omega_{ch}>\omega_{pass}$ of the
Hopf resonance
frequency \cite{bem}, which is necessary for obtaing the curve 
forms reported in Fig. 4 b under the condition of $\mu=0$, is unrealistic. 
It implies that the amplification on the BM takes place on the decaying
branch of the passive response, where viscous losses of energy are
important. Instead, we propose that the amplification
takes place 
%at 
around the passive response peak, where the group velocity
becomes small, and a large amount of energy is accumulated in the
presence of only moderate viscous forces. 
%This interpretation is
%corroborated from two-tone suppression experiments
%\cite{CM:PatuzziSHAR}.
This view is corroborated from two-tone suppression experiments,
from which the location of amplification can be determined 
\cite{CM:PatuzziSHAR}.

Our recent modeling approach \cite{AA:KernStoop2003} 
is based on $\mu<0$ and on a detailed
hydrodynamic model of the passive cochlea for the traveling wave.
Without any further modifications, this model's response yields
almost perfect agreement with experimental data for low-intensity
stimuli (see Fig. 1). The behavior of the response curves on the
high frequency branch can be brought even closer to measured data 
(see Fig. 2 (a) of \cite{AA:Magnasco2003} and Fig. 1 of 
\cite{AA:Eguiluz_etal2000}), by including the
observed mechanical forward coupling of outer hair cells 
\cite{CM:GeislerSang1995}. The
result is a perfect collapse of the response curves on
the high frequency branch \cite{AlbertDiss}. 
For stimuli of high intensity, 
additional effects evidently must be incorporated, as is indicated by the
broadened low frequency response of the experimental measurements. 
In particular, the backward coupling among outer hair cells may
become important, as by the interaction of activated Hopf amplifiers,
frequencies $\omega_i > \omega$ will be generated and back-propagated
by the coupling. When they
are actively amplified, they will lead to the observed shoulder of the
low-frequency response.
In \cite{AA:Magnasco2003}, this behavior is phenomenologically modeled 
by the filter properties, that lack the detailed relationship to cochlear 
biophysics proposed in \cite{AA:Eguiluz_etal2000}.

\begin{figure}[t]
\epsfig{file=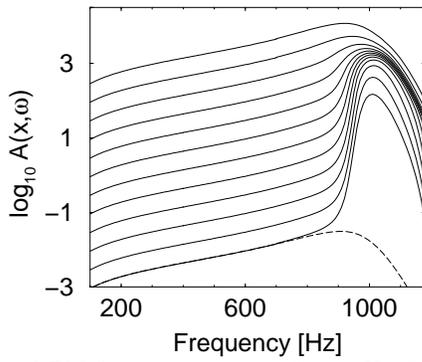, width=4.5cm, angle=-90}
\caption{
\label{LCHopfFig}
Local BM frequency response from a Hopf-type cochlea model,
where longitudinal BM couplings have been included.
Dashed line: passive response. Stimulus frequency: $\omega/2 \pi = 1000$ Hz. 
For two adjacent lines, the stimulus intensity differs by $10$ dB.
}
\end{figure}
 
\bibliography{Magnasco}

\begin{thebibliography}{9}
\bibitem[{bem()}]{bem}
\bibinfo{note}{Where {$\omega_{pass}$} is the peak frequency of the passive
  linear response, and not the resonance frequency}.
\end{thebibliography}

\end{document}